\documentclass[12pt]{amsart}
\usepackage{amssymb,latexsym}

\begin{document}
\title[Wormholes supported by phantom energy]
{Seeking exacly solvable models of traversable wormholes supported
    by phantom energy}
\author{Peter K. F. Kuhfittig}
\address{Department of Mathematics\\
Milwaukee School of Engineering\\
Milwaukee, Wisconsin 53202-3109 USA}
\date{\today}

\begin{abstract}
The purpose of this paper is to obtain exact solutions of the Einstein
field equations describing traversable wormholes supported by
phantom energy.  Their relationship to exact solutions in the
literature is also discussed, as well as the conditions required to
determine such solutions.
\end{abstract}

\maketitle

PAC number(s): 04.20.Jb, 95.36.+x

\section{Introduction}
Wormholes may be defined as handles or tunnels in the spacetime
topology linking different universes or widely separated regions
of our own universe \cite{MT88}.  Renewed interest in the subject is
due in part to the discovery that our universe is undergoing an
accelerated expansion \cite{aR98, sP99}, that is, $\overset{..}{a}
>0$ in the Friedmann equation $\overset{..}{a}/a=-\frac{4\pi}{3}
(\rho+3p)$. (Our units are taken to be those in which $G=c=1$.)
The acceleration is caused by a negative pressure \emph{dark energy}
with equation of state $p=-K\rho$, $K>\frac{1}{3}$ and $\rho>0$.
A value of $K>\frac{1}{3}$ is required for accelerated expansion;
$K=1$ corresponds to a cosmological constant \cite{mC01}.  Of
particular interest is the case $K>1$, referred to as
\emph{phantom energy}. For this case, $\rho+p<0$, in violation of
the null energy condition. (The null energy condition requires the
stress-energy tensor $T_{\alpha\beta}$ to obey
$T_{\alpha\beta}k^{\alpha}k^{\beta}\ge0 $
for all null vectors.)  It is well known that the violation of the
null energy condition is a necessary condition for the existence of
wormholes \cite{MT88}.  In this context such matter is usually called
\emph{exotic}.  Phantom energy could therefore automatically qualify
as a candidate for exotic matter, except for one problem: the notion
of dark or phantom energy applies to a homogeneous distribution of
matter in the universe, while wormhole spacetimes are necessarily
inhomogeneous.  Fortunately, the extension to spherically
symmetric inhomogeneous spacetimes has been carried out.
(See Ref.~\cite{sS05} for details.)

Returning to the definition of phantom energy, we saw that the
condition $K>1$ results in a violation of the null energy
condition. So if $K<1$, the null energy condition is met and
turns out to have a direct effect on the so-called flare-out
condition.  We shall return to this point in Section~\ref
{S:general}.

A recurring problem in the general theory of relativity is finding
exact solutions to the Einstein field equations.  In the case of
phantom-energy supported wormholes several solutions already exist
in the literature \cite{sS05, fL05, oZ05}. The strategy in this
paper is to start with a general line element, together with the
above equation of state, and to determine the conditions required
to obtain explicit exact solutions, both old and new.  Two new
solutions are discussed.

One problem is that not everyone uses the terms \emph{exact} and
\emph{explicit} in the same sense.  Thus an ``exact" solution may
contain an intractable integral, while an ``explicit" solution may
contain functions defined only implicitly.  Our primarily interest
is therefore centered on \emph{elementary functions},
rather than arbitrary functions. For present purposes these may be
defined as functions of a single variable built up by using that
variable and constants together with a finite number of algebraic
operations, composition, forming trigonometric functions and their
inverses, and constructing exponents and logarithms.

While the derivative of an elementary function is elementary, the
integral may not be.  For example, $\int e^{x^2}dx$ is not an
elementary function, or, as it is often expressed, the integral
cannot be written explicitly (or in closed form or in finite terms.)
General criteria for integration in finite terms can be found in
Refs. \cite{mR76, MZ94}.

We have similar requirements for the solution of differential
equations.  We are interested in finding solutions that can be
expressed explicitly in terms of elementary functions, as
opposed to infinite-series or numerical solutions.

The solutions in Refs.~\cite{sS05, fL05, oZ05} mentioned above
are examples of exact solutions in the sense defined here.

\section{The problem}
Consider the general line element
\begin{equation}\label{E:line1}
  ds^2=-e^{2\Phi(r)}dt^2+e^{2\alpha(r)}dr^2+r^2(d\theta^2+
   \text{sin}^2\theta\,d\phi^2),
\end{equation}
where $\Phi$ and $\alpha$ are functions of the radial coordinate $r$.
The function $\Phi$ is called the \emph{redshift function}.  We
require that $e^{2\Phi(r)}$ never be zero to avoid an event horizon.
The function $\alpha$ has a vertical asymptote at the throat $r=r_0$:
\[
   \lim_{r \to r_0+}\alpha(r)=+\infty.
\]
The reason is its relationship to the shape function $b(r)$:
\begin{equation}
   e^{2\alpha(r)}=\frac{1}{1-\frac{b(r)}{r}}.
\end{equation}
It follows that
\begin{equation}\label{E:shape}
  b(r)=r(1-e^{-2\alpha(r)}).
\end{equation}
The shape function determines the spatial shape of the wormhole when
viewed, for example, in an embedding diagram. To obtain a
traversable wormhole, the shape function must obey the usual
flare-out conditions at the throat \cite{MT88}:
$b(r_0)=r_0, b'(r_0)<1$, and $b(r)<r$.  Another requirement is
asymptotic flatness, that is, $b(r)/r\rightarrow 0$ as $r\rightarrow
 \infty$. Because of the spherical symmetry, the nonzero components
of the stress-energy tensor are $T_{00}=\rho(r)$, $T_{11}=p(r)$, and
$T_{22}=T_{33}=p_t(r)$, the transverse pressure.  The components
of the Einstein tensor in the orthonormal frame are given next
\cite{pK02}:
\begin{equation}\label{E:Einstein1}
   G_{\hat{t}\hat{t}}=\frac{2}{r}e^{-2\alpha(r)}\alpha'(r)
      +\frac{1}{r^2}(1-e^{-2\alpha(r)}),
\end{equation}
\begin{equation}\label{E:Einstein2}
   G_{\hat{r}\hat{r}}=\frac{2}{r}e^{-2\alpha(r)}\Phi'(r)
       -\frac{1}{r^2}(1-e^{-2\alpha(r)}),
\end{equation}
\begin{multline}\label{E:Einstein3}
   G_{\hat{\theta}\hat{\theta}}=G_{\hat{\phi}\hat{\phi}}\\
      =e^{-2\alpha(r)}\left(\Phi''(r)-\Phi'(r)\alpha'(r)+
      [\Phi'(r)]^2+\frac{1}{r}\Phi'(r)-\frac{1}{r}\alpha'(r)
         \right).
\end{multline}

From the Einstein field equations $G_{\hat{\alpha}\hat{\beta}}=
8\pi T_{\hat{\alpha}\hat{\beta}}$ and the equation of state
$p=-K\rho$, we have $G_{\hat{t}\hat{t}}=8\pi\rho$ and
$G_{\hat{r}\hat{r}}=8\pi(-K\rho)$, giving us the following system
of equations:
\begin{equation}\label{E:density}
  G_{\hat{t}\hat{t}}=8\pi T_{\hat{t}\hat{t}}=8\pi\rho
=\frac{2}{r} e^{-2\alpha(r)}\alpha'(r)+\frac{1}{r^2}
         (1-e^{-2\alpha(r)}),
\end{equation}
\begin{equation}\label{E:pressure}
   G_{\hat{r}\hat{r}}=8\pi T_{\hat{r}\hat{r}}=8\pi(-K\rho)
=\frac{2}{r}
      e^{-2\alpha(r)}\Phi'(r)-\frac{1}{r^2}
          (1-e^{-2\alpha(r)}).
\end{equation}
Substitution yields
\begin{multline*}
   \frac{2}{r}e^{-2\alpha(r)}\alpha'(r)+\frac{1}{r^2}
           (1-e^{-2\alpha(r)})\\
    =-\frac{1}{K}\frac{2}{r}e^{-2\alpha(r)}\Phi'(r)
       +\frac{1}{K}\frac{1}{r^2}(1-e^{-2\alpha(r)}).
\end{multline*}
After rearranging the terms,
\begin{equation}\label{E:diffeq}
   K\alpha'(r)=-\Phi'(r)-\frac{1}{2r}(e^{2\alpha(r)}-1)(K-1).
\end{equation}

This equation shows the close relationship between $\Phi'(r)$ and
$\alpha'(r)$ and hence between $\Phi(r)$ and $\alpha(r)$.  Since
$\alpha(r)\rightarrow +\infty$ as $r\rightarrow r_0+$, there is a
distressing tendency for $e^{2\Phi(r)}$ to go to zero as
$r\rightarrow r_0+$.  So, while the existence of exotic matter does
help to satisfy a basic requirement, the equation of state makes it
very difficult to obtain an exact solution without an event horizon.


\section{The redshift function}
Returning to Eq. (\ref{E:diffeq}), one way to solve this equation is to
insert the redshift function ``by hand."  One possibility is
$\Phi'(r)\equiv 0$, resulting in $\Phi=\text{constant}$.  This equation
is readily solved and leads to
\begin{equation}\label{E:Lobo}
      e^{2\alpha(r)}
     =\frac{1}{1-\left(\frac{r_0}{r}\right)^{1-1/K}}.
\end{equation}
This is the solution in Ref.~\cite{fL05}.

The only other possibility is
$\Phi(r)=\frac{1}{2}\text{ln}(r_1/r)$, for some constant $r_1$,
which allows the solution of Eq.~(\ref{E:diffeq}) by separation of
variables, that is, by factoring $1/r$.  This approach yields
\begin{equation}\label{E:Zaslavskii}
   e^{2\alpha(r)}=\frac{1}{\left(1-\frac{1}{K}\right)
     \left(1-\frac{r_0}{r}\right)}.
\end{equation}
This is the solution in Ref.~\cite{oZ05}.

It follows that to obtain an exact solution, $\Phi'(r)$ must
depend directly on $\alpha(r)$ and $\alpha'(r)$ and therefore
indirectly on the shape function. This dependence may be
expressed as $\Phi'(r)=F[\alpha(r)]\alpha'(r)$, for some
elementary function $F$.


\section{The general case}\label{S:general}
Suppose we write Eq.~(\ref{E:diffeq}) as follows:
\begin{equation}\label{E:diff2}
   -K\alpha'(r)-\frac{1}{2r}(e^{2\alpha(r)}-1)(K-1)=\Phi'(r).
\end{equation}
By the above discussion, $\Phi'(r)$ must have the form
\begin{equation}\label{E:redshift}
    \Phi'(r)=F[\alpha(r)]\alpha'(r).
\end{equation}
So by Eq.~(\ref{E:diff2}),
\begin{equation}\label{E:diff3}
  -\frac{K\alpha'(r)}{e^{2\alpha(r)}-1}-\frac{1}{2r}(K-1)=
     \frac{F[\alpha(r)]\alpha'(r)}{e^{2\alpha(r)}-1}.
\end{equation}
To obtain an exact solution, we must be able to solve
Eq.~(\ref{E:diff3}) in a closed form and to express the integral
of $\Phi'(r)$ in finite terms.  Finding $\Phi(r)$ in an exact form
does not, of course, guarantee the absence of an event horizon.
For example, if $F\equiv -1$ in Eq.~(\ref{E:redshift}), then
$\Phi(r)=-\alpha(r)+C$ and $e^{2\Phi(r)}=0$\ at the throat,
which is a typical outcome!

The simplest way to meet all these requirements is by letting
\[
    F[\alpha(r)]=-\frac{K}{e^{2\alpha(r)}}.
\]
Then, Eq.~(\ref{E:diff3}) becomes
\begin{equation*}
  -\frac{K\alpha'(r)}{e^{2\alpha(r)}-1}-\frac{1}{2r}(K-1)=
       -\frac{K\alpha'(r)}{e^{2\alpha(r)}(e^{2\alpha(r)}-1)}
\end{equation*}
with initial condition $\alpha(r_0)=+\infty$. The solution is
\[
     e^{2\alpha(r)}=\frac{1}{\text{ln}(\frac{r}{r_0})
     ^{(K-1)/K}}.
\]
Integrating $-K\alpha'(r)e^{-2\alpha(r)}$ and substituting the
expression for $e^{2\alpha(r)}$ yields
\[
   \Phi(r)=\frac{1}{2}\,\text{ln}\,\,C
     \left(\frac{r}{r_0}\right)^{K-1};
\]
$C$ is the constant of integration, which needs to be
determined from the junction conditions.  So, the line element is
\begin{equation}\label{E:line2}
  ds^2=-C\left(\frac{r}{r_0}\right)^{K-1}dt^2+\frac{1}
  {\text{ln}(\frac{r}{r_0})^{(K-1)/K}}\,dr^2
    +r^2(d\theta^2+\text{sin}^2\theta\,d\phi^2).
\end{equation}

This solution has most of the required features. For example,
since $b(r)=r(1-e^{-2\alpha(r)})$ by Eq.~(\ref{E:shape}),
we have
\[
   b(r)=r\left[1-\text{ln}\left(\frac{r}{r_0}\right)
   ^{\frac{K-1}{K}}\right].
\]
It is easily checked that $b(r_0)=r_0$, $b'(r_0)=1/K<1$,
and $b(r)<r$. (Observe that if $K<1$, then $b'(r_0)>1$ and the
flare-out condition is no longer satisfied.)

Unfortunately, the resulting spacetime is not asymptotically flat
since $b(r)$ eventually decreases.  Accordingly, the wormhole
material must be cut off at some $r=a$ and joined to an external
Schwarzschild spacetime.  A natural choice for $r=a$ is the value
for which $b(r)$ becomes a maximum.  From the critical value of
$b'(r)=0$, we get $a=r_0e^{1/(K-1)}$.  At first glance this does
not look like a large distance.  According to Ref.~\cite{rC03},
 however, $K$ is likely to be very close to unity.

Matching our interior solution to the exterior Schwarzschild
solution
\begin{equation}\label{E:Schwarzschild}
  ds^2=-\left(1-\frac{2M}{r}\right)dt^2+
    \left(1-\frac{2M}{r}\right)^{-1}dr^2+r^2(d\theta^2
       +\text{sin}^2\theta\,d\phi^2)
\end{equation}
at some $r=a$ requires continuity of the metric.  As noted in
Ref~\cite{jL03}, since the components
$g_{\hat{\theta}\hat{\theta}}$ and $g_{\hat{\phi}\hat{\phi}}$ are
already continuous due to the spherical symmetry, one needs to
impose continuity only on the remaining components at $r=a$:
\[
  g_{\hat{t}\hat{t}\text{(int)}}(a)=
      g_{\hat{t}\hat{t}\text{(ext)}}(a) \qquad \text{and} \qquad
  g_{\hat{r}\hat{r}\text{(int)}}(a)=
      g_{\hat{r}\hat{r}\text{(ext)}}(a)
\]
for the interior and exterior components, respectively.  These
requirements, in turn, imply that
\[
  \Phi_{\text{int}}(a)=\Phi_{\text{ext}}(a)\qquad \text{and}\qquad
     b_{\text{int}}(a)=b_{\text{ext}}(a).
\]
In particular,
\[
   e^{2\alpha(r)}=\frac{1}{1-\frac{b(a)}{a}}
        =\frac{1}{1-\frac{2M}{a}}.
\]
So we need to determine $M=\frac{1}{2}b(a)$, the total mass of the
wormhole for $r\le a$:
\[
  M=\frac{1}{2}b(a)=\frac{1}{2}a\left[1-
    \text{ln}\left(\frac{a}{r_0}
       \right)^{\frac{K-1}{K}}\right].
\]
Choosing $a=r_0e^{1/(K-1)}$, we get
\[
   M=\frac{1}{2}r_0e^{\frac{1}{K-1}}\left(1-\frac{1}{K}\right).
\]
Returning to $\Phi(r)$, we now have
\[
   C\left(\frac{a}{r_0}\right)^{K-1}=1-\frac{2M}{a}
\]
or
\[
    C=\frac{1-2M/a}{(a/r_0)^{K-1}}.
\]
For $a=r_0e^{1/(K-1)}$, $C=1/(Ke)$.  So the line element
becomes
\begin{equation}\label{E:line3}
  ds^2=-\frac{1}{Ke}\left(\frac{r}{r_0}\right)^{K-1}dt^2+
  \frac{1}{\text{ln}(\frac{r}{r_0})^{(K-1)/K}}
    \,dr^2+r^2(d\theta^2+\text{sin}^2\theta\,d\phi^2).
\end{equation}

While the metric is continuous on the junction surface
$r=a$, the derivatives may not be.  This behavior needs to
be taken into account when discussing the surface stresses.
The following forms, proposed by  Lobo~\cite{fL05, jL03},
are suitable for this purpose:
\[
   \sigma=-\frac{1}{4\pi a}\left(\sqrt{1-\frac{2M}{a}}-
   \sqrt{1-\frac{b(a)}{a}}\right)
\]
and
\[
   \mathcal{P}=\frac{1}{8\pi a}\left(\frac{1-\frac{M}{a}}
    {\sqrt{1-\frac{2M}{a}}}-[1+a\Phi'(a)]
       \sqrt{1-\frac{b(a)}{a}}\right).
\]
Since $b(a)=2M$, the surface stress-energy $\sigma$ is zero.  It
is readily checked that at $r=a=r_0e^{1/(K-1)}$, the surface
tangential pressure $\mathcal{P}$ is also zero.  Such a
junction surface is often referred to as a boundary surface
\cite{jL03}.

An important consideration affecting the traversability is the
proper distance $\ell(r)$ from the throat to a point away from the
throat:
\[
   \ell(r)=\int_{r_0}^{r}\frac{dr}{\sqrt{\text{ln}(\frac{r}{r_0})
    ^{(K-1)/K}}},
\]
which is finite; in fact, $\ell(r_0)=0$. Unless $K$ is extremely close
to unity, $\ell(r)$ is not going to be excessively large.  For example,
if $K=1.1$ and $r=2r_0$, then $\ell(r)\approx 7.1r_0$.

A final consideration is the time dilation near the throat. Let
$v=d\ell/d\tau$, so that $d\tau=d\ell/v$ (assuming that
$\gamma=\sqrt{1-(v/c)^2}\approx 1$.)
Since $d\ell=e^{\alpha(r)}dr$ and $d\tau=e^{\Phi(r)}dt$, we have for
any coordinate interval $\Delta t$:
\begin{equation*}
  \Delta t=\int_{t_a}^{t_b}dt=
     \int_{\ell_a}^{\ell_b}e^{-\Phi(r)}\frac{d\ell}{v}=
     \int_{r_a}^{r_b}\frac{1}{v}e^{-\Phi(r)}e^{\alpha(r)}
        dr.
\end{equation*}
Going from the throat to $r$, we get
\begin{multline*}
   \Delta t=\int_{r_0}^{r}\frac{\sqrt{Ke}}{v}\sqrt{\left(\frac
   {r_0}{r}\right)^{K-1}}\frac{dr}
    {\sqrt{\text{ln}(\frac{r}{r_0})^{(K-1)/K}}}\\
    \le \int_{r_0}^{r}\frac{\sqrt{Ke}}{v}
    \frac{dr}{\sqrt{\text{ln}(\frac{r}{r_0})^{(K-1)/K}}},
\end{multline*}
which is also well behaved near the throat.


\section{Other solutions}
Returning to Eq.~(\ref{E:redshift}), the choices for $F$ appear to be
severely limited.  At least one other possibility is
$F[\alpha(r)]=-2K/(e^{2\alpha(r)}+1)$.  So by Eq.~(\ref{E:diff2}),
\[
   -K\frac{\alpha'(r)}{e^{2\alpha(r)}-1}-\frac{1}{2r}(K-1)=
    -\frac{2K\alpha'(r)}{e^{4\alpha(r)}-1}.
\]
This equation can also be readily solved to yield
\begin{equation}\label{E:line4}
   ds^2=-C\left(\frac{r}{r_0}\right)^{2(K-1)}dt^2+
   \frac{1}{(\frac{r}{r_0})^{(K-1)/K}-1}\,dr^2+
     r^2(d\theta^2+\text{sin}^2\theta\,d\phi^2).
\end{equation}
This solution has some of the same features as the solution in the previous
section: $b(r)$ satisfies the flare-out conditions and attains a maximum
value at
\[
   a=r_0\left(\frac{2K}{2K-1}\right)^{\frac{K}{K-1}}.
\]
It is readily shown that if $r=a$ is the junction surface, then
\[
    M=r_0\frac{K-1}{2K-1}\left(\frac{2K}{2K-1}\right)^{\frac{K}{K-1}}
\]
and
\[
    C=\frac{(2K-1)^{2K-1}}{(2K)^{2K}}.
\]

The cases considered so far may very well exhaust the possibilities for
getting exact solutions without event horizons.  For example,
$F[\alpha(r)]=-1/\sqrt{e^{2\alpha(r)}-1}$ in Eq.~(\ref{E:redshift}) leads to
\[
   \Phi(r)=-\int\frac{\alpha'(r)dr}{\sqrt{e^{2\alpha(r)}-1}}=
   -\text{tan}^{-1}\sqrt{e^{2\alpha(r)}-1}+C,
\]
which is finite at $r=r_0$. So there is no event horizon.  Unfortunately,
the differential equation~(\ref{E:diff2}) leads only to an implicit
solution for $e^{\alpha(r)}$:
\begin{multline*}
    -\frac{K}{2}\left[\text{ln}(e^{2\alpha(r)}-1)-2\alpha(r)\right]
     -\frac{K-1}{2}\text{ln}\,r\\
    =-\frac{1}{2}\left(-2\,\text{tan}^{-1}\sqrt{e^{2\alpha(r)}-1}
       -\frac{2}{\sqrt{e^{2\alpha(r)}-1}}\right)+C.
\end{multline*}
This equation cannot be solved explicitly for $e^{\alpha(r)}$.

The choice $F[\alpha(r)]=1/(e^{2\alpha(r)}-1)^2$ is even worse.  Not only
do we get a strictly implicit solution, but the resulting metric has an
event horizon.  Other plausible choices, such as
\[
   F[\alpha(r)]=\frac{1}{(e^{2\alpha(r)}-1)^n}\quad \text{for} \quad n>2
\]
or
\[
   F[\alpha(r)]=\frac{1}{(e^{2\alpha(r)}-1)^{m/n}}
\]
are even more complicated and do not yield explicit solutions.

Specifying $\alpha(r)$ and solving for $\Phi(r)$ leads to similar
difficulties; judging from Eq.~(\ref{E:diff2}),
\[
    \Phi'(r)=-K\alpha'(r)-\frac{1}{2r}(e^{2\alpha(r)}-1)(K-1),
\]
if
\[
   \int_{r_0}^{r}\frac{1}{2r}(e^{2\alpha(r)}-1)(K-1)dr
\]
is finite, then
\[
    \Phi(r)=-K\alpha(r)-\int_{r_0}^{r}\frac{1}{2r}(e^{2\alpha(r)}-1)(K-1)dr
\]
is likely to lead to an event horizon at $r=r_0$.  But, as we have seen,
$\Phi(r)$ can be determined in certain special cases, such as Eqs.
(\ref{E:Lobo}) and (\ref{E:Zaslavskii}).

Another special case is discussed in Ref. \cite{sS05}: if $\rho(r)=\rho_0$,
a constant for $r_0\le r\le r_1$, then Eq.~(\ref{E:density}) becomes
\[
    8\pi\rho_0=\frac{2}{r}e^{-2\alpha(r)}\alpha'(r)+\frac{1}
      {r^2}(1-e^{-2\alpha(r)}).
\]
So
\[
   8\pi\rho_0r^2=2re^{-2\alpha(r)}\alpha'(r)+(1-e^{-2\alpha(r)})=b'(r)
\]
by Eq.~(\ref{E:shape}).  This equation yields $b(r)$ and hence $\alpha(r)$
and $\Phi(r)$, although the details are complicated.


\section{conclusion}
It is shown in this paper that for wormholes supported by phantom energy
the only specific choices for the redshift function $\Phi$ that lead to
explicit exact solutions without an event horizon are
$\Phi(r)\equiv\text{constant}$ and
$\Phi(r)=\frac{1}{2}\text{ln}(r_1/r)$.  Otherwise $\Phi(r)$ must depend
on $\alpha(r)$ in such a way that $\Phi'(r)=F[\alpha(r)]\alpha'(r)$,
for some elementary function $F$.  The choices for $F$ are severely
limited.  Two new solutions are obtained.

\end{document}